\documentclass{article}
\usepackage{spconfa4,amsmath,graphicx}
\usepackage{amssymb}
\usepackage{booktabs}
\usepackage{multirow}
\usepackage{siunitx}
\usepackage{cite}
\usepackage{url}
\usepackage{xcolor}
\usepackage{tikz}
\usepackage{balance}
\usepackage{graphicx}
\usepackage{xcolor}
\usetikzlibrary{arrows.meta,positioning,fit}
\title{An Efficient Parametric Codec for Low-Bitrate First-Order Ambisonics}
\name{\fontsize{11}{13}\selectfont Wei-Ting Lai, Amy Bastine, Lachlan Birnie, Thushara D. Abhayapala, Prasanga N. Samarasinghe}
\address{\fontsize{11}{13}\selectfont\text{Audio \& Acoustic Signal Processing Group, The Australian National University, Canberra, Australia}}

\begin{document}
\ninept

\setlength{\abovedisplayskip}{4pt plus 1pt minus 1pt}
\setlength{\belowdisplayskip}{4pt plus 1pt minus 1pt}
\setlength{\abovedisplayshortskip}{2pt plus 1pt minus 1pt}
\setlength{\belowdisplayshortskip}{2pt plus 1pt minus 1pt}
\setlength{\jot}{2pt}
\setlength{\textfloatsep}{6pt plus 1pt minus 1pt}
\setlength{\floatsep}{5pt plus 1pt minus 1pt}
\setlength{\intextsep}{5pt plus 1pt minus 1pt}
\setlength{\dbltextfloatsep}{6pt plus 1pt minus 1pt}
\setlength{\dblfloatsep}{5pt plus 1pt minus 1pt}
\makeatletter
\long\def\@makecaption#1#2{%
 \vskip 3pt
 \setbox\@tempboxa\hbox{#1. #2}%
 \ifdim \wd\@tempboxa >\hsize #1. #2\par
 \else \hbox to\hsize{\hfil\box\@tempboxa\hfil}\fi}
\renewenvironment{thebibliography}[1]
 {\section{References}%
  \list{\scriptsize[\arabic{enumi}]}{%
   \settowidth\labelwidth{\scriptsize[#1]}%
   \leftmargin\labelwidth
   \advance\leftmargin\labelsep
   \usecounter{enumi}%
   \scriptsize
   \setlength{\itemsep}{0pt}%
   \setlength{\parsep}{0pt}%
   \setlength{\topsep}{1pt}%
   \setlength{\partopsep}{0pt}}%
  \def\newblock{\hskip .11em plus .33em minus .07em}%
  \sloppy\clubpenalty4000\widowpenalty4000
  \sfcode`\.=1000\relax}
 {\endlist}
\makeatother

\maketitle

\begin{abstract}
Driven by the rapid growth of immersive teleconferencing and generative spatial audio, efficient low-bitrate coding of first-order Ambisonics (FOA) has become increasingly important.
In this work, we develop a lightweight parametric FOA codec that retains the standard Directional Audio Coding analysis and synthesis while redesigning the spatial metadata quantization scheme.
Rather than quantizing direction-of-arrival (DOA) and diffuseness independently, we combine them into a 3-D directivity vector and jointly quantize these vectors across frequency bands via residual vector quantization (RVQ).
The RVQ codebooks are optimized within minutes via stage-wise $k$-means without backpropagation, yielding a constant-bitrate (CBR) representation that decouples metadata rate from the number of frequency bands.
Evaluations show that our approach outperforms low-bitrate perceptual codecs in FOA reconstruction, remains competitive with neural codecs on the downstream Sound Event Localization and Detection task, and maintains robust performance when paired with different external monaural codecs.
Given its lightweight training, strong performance, and CBR design, we consider the proposed method to be a favorable and reproducible baseline for future FOA codec research.
\end{abstract}


\vspace{-0.2cm}
\section{Introduction}
\label{sec:intro}

Recently, low-bitrate audio codecs/tokenizers have become increasingly important~\cite{soundstream, encodec, dac}, driven by growing demands in bandwidth-constrained communication, such as satellite communication~\cite{satellite}, and generative audio modeling~\cite{audiolm, musicgen}.
Audio codecs aim to compress audio signals into compact discrete representations.
Existing lossy codecs can be broadly categorized into perceptual and neural codecs.
Perceptual codecs rely on psychoacoustic models to discard perceptually less important information~\cite{mp3, opus}, whereas neural codecs learn compact representations directly from large-scale audio data~\cite{soundstream, encodec, dac}.
In recent years, neural codecs have increasingly replaced conventional perceptual codecs, offering better perceptual quality at low bitrates through more efficient learned representations.

Meanwhile, immersive teleconferencing~\cite{opusfoa} and spatial audio generation~\cite{mcsimclr, elsa, phasecoder, spatialomni} have increased the need for multichannel and spatial audio codecs~\cite{spatialcodec, banc, foatokenizer, vcnac}.
First-order Ambisonics (FOA) provides an efficient spatial audio representation through a four-channel format that supports rendering to different loudspeaker and headphone configurations~\cite{ambisonics, sh-pw, mpegh}.
While neural codecs for monaural and stereo audio are well-established, such as EnCodec~\cite{encodec} and DAC~\cite{dac}, neural FOA coding remains relatively underexplored.
However, independently applying monaural neural codecs to each FOA channel can degrade spatial fidelity~\cite{foatokenizer}.
FOA-VQGAN~\cite{foatokenizer} is one of the few recent methods that directly learns discrete tokens from FOA waveforms, whereas existing FOA codecs, such as Opus~\cite{opus, opusfoa}, MPEG-H~\cite{mpegh}, and Directional Audio Coding (DirAC)~\cite{dirac_1, dirac_2, dirac_codec1, dirac_codec2, dirac_codec3, fo-dirac} use perceptual approaches~\cite{ivas}.

For FOA coding at total bitrates below $10$~kbps, the field remains relatively underexplored, with limited state-of-the-art methods to serve as baselines.
Opus is a possible baseline~\cite{opus, opusfoa}; however, its fixed spatial preprocessing followed by mono or stereo coding can limit and substantially degrade spatial fidelity at low bitrates~\cite{fo-dirac, foatokenizer}, making it a non-ideal choice.
Similarly, MPEG-H provides strong spatial reconstruction quality, but typically operates at bitrates above the low-bitrate range considered here, making it more suitable as a high-quality reference than a directly comparable baseline.


DirAC~\cite{dirac_1, dirac_2, dirac_codec1, dirac_codec2, dirac_codec3, fo-dirac} provides a parametric alternative that decomposes FOA into audio signals and spatial parameters.
The audio signals are coded by a signal codec, for example by coding the omnidirectional channel with an external monaural codec.
The spatial parameters, including direction-of-arrival (DOA) and diffuseness, are separately transmitted by a metadata codec.
Existing DirAC metadata codecs rely on perceptual rules to determine frequency grouping, parameter quantization resolution, and temporal update rates~\cite{dirac_codec1, psychoacoustics, dirac_codec2, dirac_codec3, fo-dirac}.
To improve coding efficiency, recent implementations use variable-bitrate (VBR) metadata coding, where the transmitted bitrate varies according to the spatial characteristics of the audio content~\cite{dirac_codec2, fo-dirac}.
However, such heuristic designs complicate bit allocation and reduce reproducibility as a baseline.
This motivates a more efficient and reproducible FOA codec with constant-bitrate (CBR) DirAC spatial metadata.

In this work, we propose a lightweight parametric FOA codec that introduces an efficient CBR spatial metadata quantizer while retaining the rest of the standard DirAC pipeline.
We combine DOA and diffuseness parameters into a 3-D directivity vector and jointly quantize them across bands using a residual vector quantizer (RVQ)~\cite{rvq,soundstream}, where the codebooks are fitted offline via $k$-means~\cite{k-means} within minutes.
Results show that the proposed method maintains robust performance for both FOA reconstruction and downstream Sound Event Localization and Detection (SELD)~\cite{starss22, starss23} when paired with different monaural neural signal codecs.
We consider the proposed method to be a favorable baseline for future FOA codec comparisons, as it supports efficient CBR configurations across different target rates.

\vspace{-0.3cm}
\section{Problem Formulation}
\label{sec:problem}

Let the FOA signal in ACN/SN3D format at time frame $t\in\{1,\ldots,T\}$ and frequency bin $k\in\{1,\ldots,K\}$ be $\mathbf{a}_{t,k}=[W_{t,k},\\Y_{t,k},Z_{t,k},X_{t,k}]^{\mathsf T}\in\mathbb{C}^{4}$.

DirAC provides a parametric representation of FOA through an analysis stage and a synthesis stage~\cite{dirac_1, fo-dirac}.
The analysis stage disentangles the FOA soundfield $\mathbf{a}_{t,k}$ into an omnidirectional component $W_{t,k}$ and bandwise spatial metadata over $B$ equivalent rectangular bandwidth (ERB) frequency bands~\cite{erb}.
For each band $b\in\{1,\ldots,B\}$ at frame $t$, the bandwise intensity vector $\mathbf{I}_{t,b}$ and sound-field energy $E_{t,b}$ are computed as
\begin{align}
    \mathbf{I}_{t,b}
    &= \textstyle\sum_{k\in\mathcal{K}_b}
    \operatorname{Re}\!\left\{W_{t,k}\mathbf{U}_{t,k}^{*}\right\},
    \label{eq:intensity}\\
    E_{t,b}
    &= \tfrac{1}{2}\textstyle\sum_{k\in\mathcal{K}_b}
    \left(|W_{t,k}|^{2}+\|\mathbf{U}_{t,k}\|_{2}^{2}\right),
    \label{eq:energy}
\end{align}
where $\mathcal{K}_b$ denotes the STFT bins in band $b$ and $\mathbf{U}_{t,k}=[Y_{t,k},Z_{t,k},\\X_{t,k}]^{\mathsf T}$ denotes the first-order spherical harmonic (SH) components.

From $\mathbf{I}_{t,b}$ and $E_{t,b}$, the DOA $\boldsymbol{\Omega}_{t,b}\in\mathbb{S}^{2}$ and diffuseness $D_{t,b}\in[0,1]$ are derived as
\begin{equation}
    \boldsymbol{\Omega}_{t,b}
    =
    -\frac{\mathbf{I}_{t,b}}{\|\mathbf{I}_{t,b}\|_{2}},
    \qquad
    D_{t,b}
    =
    1-\frac{\|\mathbf{I}_{t,b}\|_{2}}{E_{t,b}}.
    \label{eq:dirac_params}
\end{equation}

\begin{figure}[t]
    \centering
    \includegraphics[width=\linewidth]{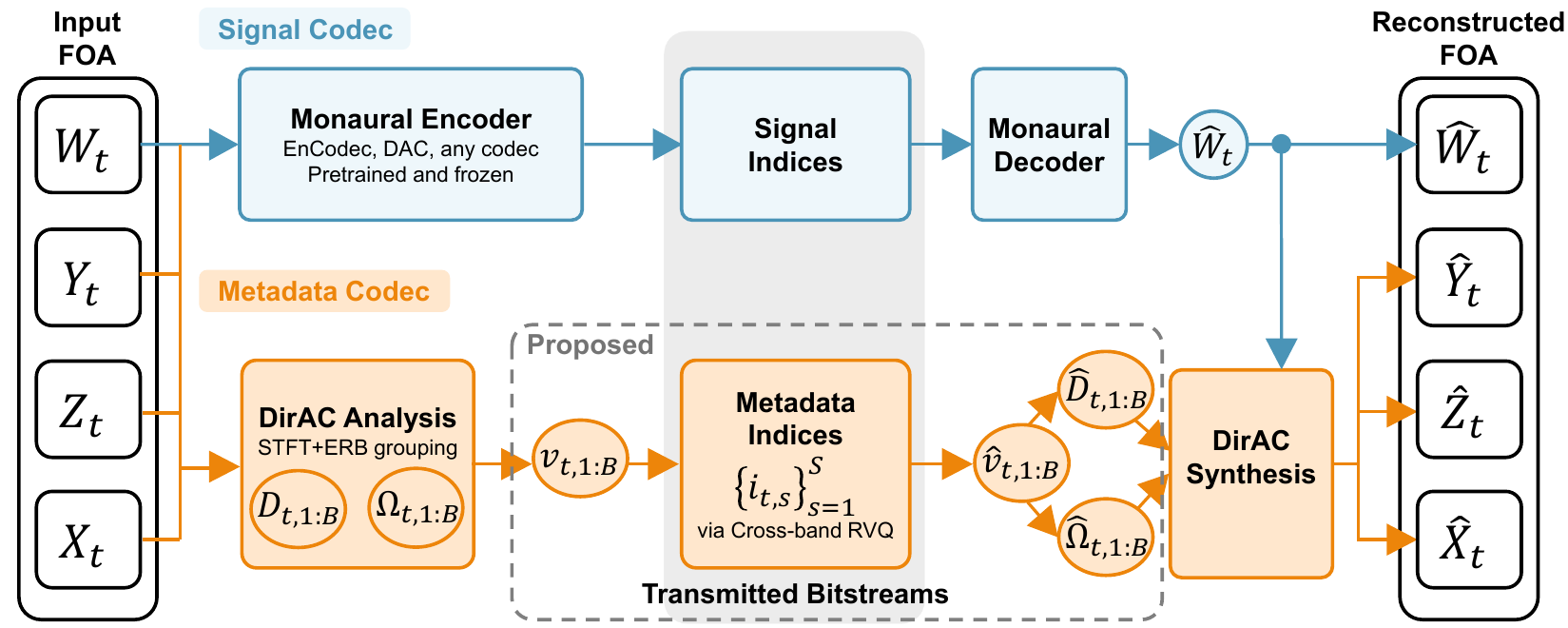}
    \caption{Overview of the proposed parametric FOA codec. 
    }
    \label{fig:problem}
    \vspace{-0.2cm}
\end{figure}
Collecting these continuous-valued spatial parameters across all $B$ bands yields the spatial metadata for frame $t$, denoted by $\mathcal{M}_{t} = \{(\boldsymbol{\Omega}_{t,b}, D_{t,b})\}_{b=1}^{B}$.

The omnidirectional channel $W_{t,k}$ and spatial metadata $\mathcal{M}_{t}$ are encoded independently.
An external monaural signal codec encodes and decodes $W_{t,k}$ to yield the reconstructed omnidirectional signal $\hat{W}_{t,k}$, whereas a spatial metadata quantizer $Q(\cdot)$ maps the continuous metadata $\mathcal{M}_{t}$ to a set of discrete quantization indices $\mathbf{z}_{t}$, from which the reconstructed metadata $\hat{\mathcal{M}}_{t}$ is obtained via the dequantization mapping $Q^{-1}(\cdot)$ as
\begin{equation}
    \mathbf{z}_{t} = Q(\mathcal{M}_{t}), 
    \qquad 
    \hat{\mathcal{M}}_{t} = Q^{-1}(\mathbf{z}_{t}) = \{(\hat{\boldsymbol{\Omega}}_{t,b}, \hat{D}_{t,b})\}_{b=1}^{B}.
    \label{eq:quantization}
\end{equation}

The DirAC synthesis then uses the dequantized spatial parameters $\hat{\mathcal{M}}_{t}$ together with the decoded omnidirectional signal $\hat{W}_{t,k}$ to synthesize the first-order SH channels. The complete reconstructed FOA signal is formed as
\begin{equation}
\hat{\mathbf a}_{t,k}
=
\begin{bmatrix}
\hat W_{t,k}\\
\hat{\mathbf {U}}_{t,k}
\end{bmatrix},
    \label{eq:reconstructed_foa}
\end{equation}
where the omnidirectional component $\hat{W}_{t,k}$ is directly preserved, and the spatial channels $\hat{\mathbf{U}}_{t,k} = [\hat{Y}_{t,k}, \hat{Z}_{t,k}, \hat{X}_{t,k}]^{\mathsf T}$ are synthesized by superimposing directional component $\hat{\mathbf{U}}_{t,k}^{\mathrm{dir}}$ and diffuse component $\hat{\mathbf{U}}_{t,k}^{\mathrm{diff}}$ as
\begin{equation}
    \hat{\mathbf{U}}_{t,k} = \hat{\mathbf{U}}_{t,k}^{\mathrm{dir}} + \hat{\mathbf{U}}_{t,k}^{\mathrm{diff}}.
    \label{eq:synthesis_total}
\end{equation}

The directional branch synthesizes spatial components corresponding to a plane wave arriving from $\hat{\boldsymbol{\Omega}}_{t,b}$ as
\begin{equation}
    \hat{\mathbf{U}}_{t,k}^{\mathrm{dir}} = \sqrt{1 - \hat{D}_{t,b}} \, \hat{W}_{t,k} \, \mathbf{y}_{1}(\hat{\boldsymbol{\Omega}}_{t,b}),
    \label{eq:synthesis_dir}
\end{equation}
where $\mathbf{y}_{1}(\hat{\boldsymbol{\Omega}}_{t,b}) = [Y_1^{-1}(\hat{\boldsymbol{\Omega}}_{t,b}), Y_1^{0}(\hat{\boldsymbol{\Omega}}_{t,b}), Y_1^{1}(\hat{\boldsymbol{\Omega}}_{t,b})]^{\mathsf T} \in \mathbb{R}^3$ represents the first-order real SH basis vector.

The diffuse branch synthesizes an isotropic diffuse field from mutually decorrelated versions of $\hat{W}_{t,k}$ as
\begin{equation}
    \hat{\mathbf{U}}_{t,k}^{\mathrm{diff}} = \sqrt{\hat{D}_{t,b}} \, \mathbf{d}_{t,k}(\hat{W}_{t,k}),
    \label{eq:synthesis_diff}
\end{equation}
where $\mathbf{d}_{t,k}(\cdot)\in\mathbb{C}^{3}$ is the DirAC diffuse renderer that generates mutually decorrelated signals to simulate ambient reverberation~\cite{dirac_1}.

Conventional DirAC metadata codecs quantize DOA and diffuseness separately using perceptually designed parameter resolutions~\cite{dirac_codec1, fo-dirac}. 
Specifically, $D_{t,b}$ is mapped to a non-uniform codebook $\mathcal{C}_D$, while $\boldsymbol{\Omega}_{t,b}$ is quantized onto a spherical grid $\mathcal{G}_\Omega$ whose angular resolution is adapted according to diffuseness. 
The number of parameter bands $B$ is also reduced by merging adjacent frequency bands.
In low-bitrate transmission scenarios, however, these heuristic psychoacoustic designs of $\mathcal{C}_D$, $\mathcal{G}_\Omega$, and $B$ lead to VBR transmission and inefficient parameter allocation. 
The goal of this paper is therefore to develop an efficient CBR DirAC metadata quantizer $Q(\cdot)$ that improves the reconstruction fidelity of $\hat{\mathbf{a}}_{t,k}$ under strict low-bitrate constraints.

\vspace{-0.4cm}
\section{Proposed Method}
\label{sec:method}

\begin{table*}[t]
\centering
\caption{
FOA reconstruction performance on SpatialVCTK, STARSS23, and MEIR.
Lower is better for all metrics.
}
\vspace{0.05cm}
\label{tab:main_results}

\scriptsize
\setlength{\tabcolsep}{1.5pt}
\renewcommand{\arraystretch}{1.00}

\begin{tabular*}{\textwidth}{
@{\extracolsep{\fill}}
llccc|
ccccc|
cccc|
cccc
@{}
}
\toprule

\multirow{2}{*}{\textbf{Method}}
& \multirow{2}{*}{\textbf{W-ch}}
& \multirow{2}{*}{$\mathbf{R_{\rm sp}}$}
& \multirow{2}{*}{$\mathbf{S}$}
& \multirow{2}{*}{$\mathbf{R_{\rm tot}}$}
& \multicolumn{5}{c|}{\textbf{SpatialVCTK}}
& \multicolumn{4}{c|}{\textbf{STARSS23}}
& \multicolumn{4}{c}{\textbf{MEIR}} \\

\cmidrule(lr){6-10}
\cmidrule(lr){11-14}
\cmidrule(lr){15-18}

& & & & &
\textbf{STFT}
& \textbf{Mel}
& \textbf{Ang}
& \textbf{Diff}
& \textbf{WER}
&
\textbf{STFT}
& \textbf{Mel}
& \textbf{Ang}
& \textbf{Diff}
&
\textbf{STFT}
& \textbf{Mel}
& \textbf{Ang}
& \textbf{Diff} \\
\midrule

MPEG-H~\cite{mpegh}
& --
& -- & -- & $\approx$48
& 1.31 & 0.82 & \textbf{2.11}$^\circ$ & 0.02 & 0.20
& 1.45 & \textbf{0.81} & \textbf{10.37}$^\circ$ & \textbf{0.05}
& 2.39 & 1.47 & 27.53$^\circ$ & \textbf{0.16} \\

\specialrule{0.7pt}{0.8pt}{0.3pt}

FOA-VQGAN~\cite{foatokenizer}
& --
& -- & -- & 0.9
& 1.62 & 1.39 & 3.96$^\circ$ & -- & 0.67
& -- & -- & -- & --
& 1.82 & 1.43 & 25.83$^\circ$ & -- \\

\specialrule{0.7pt}{0.8pt}{0.3pt}

\multirow{2}{*}{Opus~\cite{opus}}
& --
& -- & -- & $\approx$24
& 2.42 & 1.70 & 10.04$^\circ$ & 0.11 & 0.23
& 2.50 & 1.73 & 28.46$^\circ$ & 0.12
& 4.16 & 2.48 & 56.53$^\circ$ & 0.27 \\

& --
& -- & -- & $\approx$32
& 2.22 & 1.44 & 5.31$^\circ$ & 0.05 & 0.19
& 2.35 & 1.52 & 19.18$^\circ$ & 0.08
& 4.02 & 2.27 & 38.67$^\circ$ & 0.24 \\

\specialrule{0.7pt}{0.8pt}{0.3pt}

\multirow{4}{*}{DirAC~\cite{fo-dirac}}
& EnCodec~\cite{encodec}
& 0.75 & -- & 6.75
& 2.06 & 2.21 & 29.21$^\circ$ & \underline{\textbf{0.00}} & 0.15
& 2.04 & 1.67 & 47.42$^\circ$ & 0.30
& 2.24 & 1.45 & 31.22$^\circ$ & 0.24 \\

& EnCodec
& 1.5 & -- & 7.5
& 1.18 & 0.94 & 3.56$^\circ$ & \textbf{0.00} & 0.15
& 1.92 & 1.60 & 39.19$^\circ$ & 0.32
& 1.97 & 1.31 & 18.24$^\circ$ & 0.25 \\

\cmidrule(lr){2-18}

& DAC~\cite{dac}
& 0.75 & -- & 6.75
& 1.91 & 2.02 & 29.21$^\circ$ & \underline{\textbf{0.00}} & \textbf{0.14}
& 1.66 & 1.22 & 47.48$^\circ$ & 0.30
& 2.17 & 1.42 & 31.16$^\circ$ & 0.24 \\

& DAC
& 1.5 & -- & 7.50
& \textbf{0.91} & \textbf{0.60} & 3.56$^\circ$ & \textbf{0.00} & \textbf{0.14}
& 1.53 & 1.13 & 39.19$^\circ$ & 0.32
& 1.87 & 1.28 & \textbf{18.15}$^\circ$ & 0.25 \\

\specialrule{0.7pt}{0.8pt}{0.3pt}
\specialrule{0.7pt}{0.3pt}{0.8pt}

\multirow{6}{*}{Proposed}
& EnCodec
& 0.75 & 5 & 6.75
& 1.23 & 0.97 & 3.51$^\circ$ & 0.02 & 0.15
& 1.65 & 1.44 & 37.98$^\circ$ & \underline{0.20}
& 1.78 & 1.24 & 20.84$^\circ$ & 0.22 \\

& EnCodec
& 1.5 & 10 & 7.5
& 1.22 & 0.95 & 2.72$^\circ$ & 0.02 & 0.15
& 1.65 & 1.43 & 34.89$^\circ$ & 0.20
& 1.78 & 1.24 & 19.61$^\circ$ & 0.22 \\

& EnCodec
& 3.0 & 20 & 9
& 1.21 & 0.94 & 2.18$^\circ$ & 0.02 & 0.15
& 1.64 & 1.43 & 33.06$^\circ$ & 0.20
& 1.78 & 1.24 & 18.84$^\circ$ & 0.23 \\

\cmidrule(lr){2-18}

& DAC
& 0.75 & 5 & 6.75
& \underline{0.99} & \underline{0.67} & \underline{3.51$^\circ$} & 0.02 & \underline{\textbf{0.14}}
& \underline{1.26} & \underline{0.96} & \underline{37.80$^\circ$} & 0.20
& \underline{1.69} & \underline{1.21} & \underline{20.76$^\circ$} & \underline{0.22} \\

& DAC
& 1.5 & 10 & 7.5
& 0.97 & 0.65 & 2.72$^\circ$ & 0.02 & \textbf{0.14}
& 1.26 & 0.96 & 34.61$^\circ$ & 0.20
& 1.68 & 1.21 & 19.51$^\circ$ & 0.22 \\

& DAC
& 3.0 & 20 & 9.0
& 0.96 & 0.63 & 2.17$^\circ$ & 0.02 & \textbf{0.14}
& \textbf{1.25} & 0.95 & 32.73$^\circ$ & 0.20
& \textbf{1.68} & \textbf{1.20} & 18.75$^\circ$ & 0.23 \\

\bottomrule
\end{tabular*}

\vspace{0.05cm}
\raggedright
\tiny
$W$-ch denotes the external monaural signal codec used to encode and decode the omnidirectional $W$ channel. $R_{\rm sp}$ and $R_{\rm tot}$ denote the nominal spatial-metadata and total bitrates in kbps, respectively. $R_{\rm tot}$ values for Opus and MPEG-H are approximate due to their VBR operation. Ang and Diff denote angular and diffuseness errors, respectively, and WER is reported in percent. The best result per metric is shown in \textbf{bold}, and the best result per metric at $R_{\rm sp} = 0.75$~kbps is \underline{underlined}.
\end{table*}

We propose a lightweight FOA codec that retains the standard DirAC analysis and synthesis~\cite{fo-dirac}, redesigning only the spatial metadata quantizer $Q(\cdot)$. 
Specifically, spatial parameters are represented as 3-D directivity vectors and jointly quantized across bands via RVQ~\cite{rvq, soundstream}, as illustrated in Fig.~\ref{fig:problem}.

\vspace{-0.2cm}
\subsection{3-D Directivity Vector}
Unlike conventional DirAC metadata coding, which quantizes DOA and diffuseness separately, we combine them into a 3-D directivity vector based on the directional scaling in~\eqref{eq:synthesis_dir}.
Specifically, we define
\begin{equation}
    \mathbf{v}_{t,b}
    =
    \sqrt{1-D_{t,b}}\,
    \boldsymbol{\Omega}_{t,b}.
    \label{eq:directivity}
\end{equation}

The direction of $\mathbf{v}_{t,b}$ represents the DOA, while its norm corresponds directly to the directional gain used in DirAC synthesis.
Thus, as diffuseness increases, $\|\mathbf{v}_{t,b}\|_2$ decreases and directional errors are naturally downweighted during vector quantization, without requiring an explicit diffuseness-dependent DOA resolution.

At the decoder, the DirAC parameters are recovered from the quantized vector as
\begin{equation}
    \hat{D}_{t,b}
    =
    1-\|\hat{\mathbf{v}}_{t,b}\|_2^2,
    \qquad
    \hat{\boldsymbol{\Omega}}_{t,b}
    =
    \frac{\hat{\mathbf{v}}_{t,b}}
    {\|\hat{\mathbf{v}}_{t,b}\|_2}.
    \label{eq:directivity_inverse}
\end{equation}

When $\|\hat{\mathbf{v}}_{t,b}\|_2=0$, the DOA is irrelevant to synthesis and can be assigned an arbitrary unit direction, e.g., $\hat{\boldsymbol{\Omega}}_{t,b}=[1,0,0]^{\mathsf T}$.

\vspace{-0.2cm}

\subsection{Cross-Band Residual Vector Quantization}

Encoding each band vector $\mathbf{v}_{t,b}$ independently requires separate bit allocation across bands.
Instead, we stack all $B$ band vectors into a multiband directivity matrix
$\mathbf{V}_{t}=[\mathbf{v}_{t,1},\dots,\mathbf{v}_{t,B}]^{\mathsf T}
\in\mathbb{R}^{B\times3}$
and quantize $\mathbf{V}_{t}$ jointly using cross-band RVQ~\cite{rvq, soundstream}.

The quantizer consists of $S$ stages.
For stage $s\in\{1,\ldots,S\}$, let
$\mathcal{C}_{s}=\{\mathbf{C}_{s,m}\}_{m=0}^{M-1}
\subset\mathbb{R}^{B\times3}$
denote a pre-fitted codebook of $M$ cross-band correction patterns, with
$\mathbf{C}_{s,0}=\mathbf{0}$ reserved as an idle codeword.

Starting from $\hat{\mathbf{V}}_{t}^{(0)}=\mathbf{0}$, each RVQ stage tests every codeword by adding it to the reconstruction from the previous stage. We denote the resulting matrix at stage $s$ as $\tilde{\mathbf{V}}_{t,m}^{(s)}=\hat{\mathbf{V}}_{t}^{(s-1)}+\mathbf{C}_{s,m}$.

To ensure that each reconstructed band remains within the unit sphere, we define the row-wise projection as
\begin{equation}
    \bigl[\Pi(\tilde{\mathbf{V}}_{t,m}^{(s)})\bigr]_{b,:}
    =
    \frac{ \bigl[\tilde{\mathbf{V}}_{t,m}^{(s)})\bigr]_{b,:}}
    {\max\bigl(1,\lVert\bigl[\tilde{\mathbf{V}}_{t,m}^{(s)})\bigr]_{b,:}\rVert_{2}\bigr)}.
    \label{eq:projection}
\end{equation}

Let $e_t=\sum_{b=1}^{B}E_{t,b}$ denote the total energy of frame $t$.
For $e_t>0$, the distortion of each estimated matrix is measured as
\begin{equation}
d_t\!\left(
\mathbf{V}_t,
\Pi(\tilde{\mathbf{V}}_{t,m}^{(s)})
\right)
=
\frac{
\sum_{b=1}^{B} E_{t,b}
\left\|
[\mathbf{V}_t]_{b,:}
-
[\Pi(\tilde{\mathbf{V}}_{t,m}^{(s)})]_{b,:}
\right\|_2^2
}{
e_t
},
\label{eq:distortion}
\end{equation}
where frames with $e_t=0$ are ignored.

Each RVQ stage greedily selects the codeword with the minimum distortion as
\begin{equation}
    i_{t,s}
    =
    \arg\min_{m}
    d_{t}
    \left(
        \mathbf{V}_{t},
        \Pi(\tilde{\mathbf{V}}_{t,m}^{(s)})
    \right),
\end{equation}
and the reconstruction is updated as
\begin{equation}
    \hat{\mathbf{V}}_{t}^{(s)} = \Pi(\tilde{\mathbf{V}}_{t,i_{t,s}}^{(s)}).
\end{equation}

After $S$ stages, the final quantized output is $\hat{\mathbf{V}}_{t}=\hat{\mathbf{V}}_{t}^{(S)}$, whose $b^\text{th}$ row is $\hat{\mathbf{v}}_{t,b}$.

\vspace{-0.2cm}
\subsection{Codebook Fitting and CBR Operation}
The RVQ codebooks are fitted offline using stage-wise $k$-means~\cite{k-means}. 
For each stage $s$, the codebook $\mathcal{C}_s$ is fitted based on the reconstruction from the previous stage $\hat{\mathbf{V}}_{t}^{(s-1)}$ to minimize the total energy-weighted distortion $\sum_t e_t\, d_t\!\left(\mathbf{V}_t,\hat{\mathbf{V}}_t^{(s)}\right)$.


Once fitted, the codebooks remain fixed and directly compress the spatial metadata of arbitrary input FOA signals without further optimization.
With a frame update period of $T_{\mathrm{u}}$, the quantizer outputs $S$ discrete indices $\{i_{t,s}\}_{s=1}^S$ per frame $t$, yielding a constant spatial metadata bitrate $R_{\mathrm{sp}} = (S\log_{2}M) / T_{\mathrm{u}}$.
Because $R_{\mathrm{sp}}$ is completely decoupled from the number of parameter bands $B$, the system can employ a fine frequency resolution without increasing the metadata bitrate budget.

\begin{figure*}[t]
    \centering
    \includegraphics[width=\textwidth]{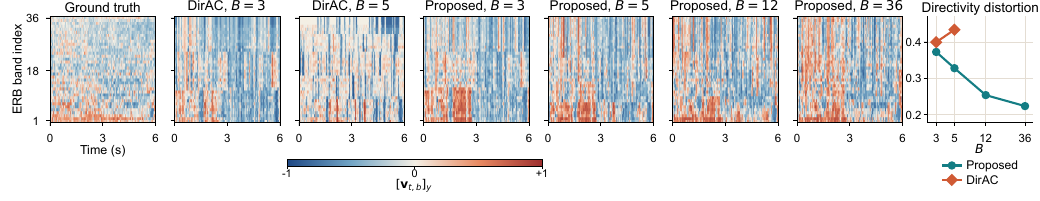}
    \vspace{-0.7cm}
    \caption{Reconstruction of the 3-D directivity vector for a STARSS23 clip ($R_{\mathrm{sp}}=750$~bps for both methods). For visualization, we show the $y$-axis component $[\mathbf{v}_{t,b}]_y$. The rightmost curve reports the energy-weighted distortion $\mathcal{D}=\sum_t e_t d_t/\sum_t e_t$.}
    \label{fig:varyB}
\end{figure*}

\vspace{-0.2cm}
\section{Experimental Validation}
\label{sec:experiments}

\vspace{-0.1cm}
\subsection{Experimental Setup}
This section describes the experimental setup for evaluating the proposed method on FOA reconstruction and downstream SELD, including codebook fitting, datasets, baselines, and evaluation metrics.

For offline codebook fitting of the proposed quantizer, we generate a simulated FOA dataset using the image-source method in Pyroomacoustics~\cite{pyroomacoustics, ism}. Dry source signals are drawn from LibriSpeech (speech)~\cite{librispeech}, ESC-50 (general sounds)~\cite{esc}, and ASE1K (transients)~\cite{ase}. A total of 120,000 6-second ACN/SN3D FOA scenes are synthesized at 24~kHz ($\approx200$~h), with reverberation times $T_{60} \in [0, 1.2]$~s, random source gains in $[-6, 6]$~dB, and additive white Gaussian noise at an SNR of 30~dB. Most scenes contain 1--3 static sources, while 12,000 scenes contain 10 concurrent sources to simulate dense cocktail-party scenarios.

To extract DirAC metadata, a 1024-point STFT with a Hann window and a 120-sample hop size is applied, grouping frequency bins into $B=36$ parameter bands across 0--12~kHz based on ERB~\cite{erb}.
The extracted metadata are temporally averaged to a fixed update period of $T_{\mathrm{u}}=40$~ms.
To reduce computational cost, we sample 8 out of the 150 frames from each clip. 
After discarding zero-energy silent frames, a total of $T = 932{,}102$ frames are retained for codebook fitting. 
Unless otherwise stated, we use $S=5$ RVQ stages with $M=64$ codewords per stage as the default configuration, resulting in $R_{\mathrm{sp}}=0.75$~kbps.
At each RVQ stage, $k$-means is run for 18 iterations from three deterministic initializations, and the solution with the lowest distortion is retained. 
The complete codebook fitting process takes approximately $7$~min on an NVIDIA RTX 3070 GPU.

Reconstruction performance is evaluated on SpatialVCTK, created by spatializing VCTK speech recordings under free-field conditions~\cite{vctk}, STARSS23~\cite{starss23}, which contains real-world FOA recordings, and FOA-MEIR~\cite{meir}, constructed using real measured room impulse responses and spatial background noise.

We compare against four baseline codecs. MPEG-H 3D Audio~\cite{mpegh} is included as a high-bitrate reference for FOA coding. For conventional DirAC, we reproduce the coding scheme in~\cite{fo-dirac}. Since the original work does not fully specify the bitrate-control details, we implement a CBR variant following its key principle that fewer DOA bits are allocated as diffuseness increases. FOA-VQGAN is included as a representative end-to-end neural FOA codec. As its implementation is not publicly available, we report the published results~\cite{foatokenizer} under matched evaluation conditions where applicable. We additionally evaluate Opus using four-channel mapping family 1~\cite{opus}, constrained VBR, and a 10-ms frame duration. 

To examine the framework's robustness when paired with different monaural signal codecs, the $W$ channel is encoded using pretrained EnCodec~\cite{encodec} or DAC~\cite{dac} at 6~kbps for both DirAC and the proposed method. 
The monaural codecs are kept fixed, and their decoded $\hat{W}$ outputs are directly used for DirAC synthesis.
We additionally evaluate the impact of varying the number of RVQ stages $S$ and the number of parameter bands $B$, with smaller $B$ obtained by pooling adjacent ERB bands, to analyze the rate--distortion trade-off and verify whether preserving finer frequency resolution under a decoupled constant bitrate benefits spatial reconstruction.

FOA reconstruction quality is evaluated using STFT and Mel-spectral losses implemented with Auraloss~\cite{auraloss}.
For SpatialVCTK, we additionally evaluate Word Error Rate (WER) by transcribing the reconstructed $\hat{W}$ channel using the pretrained multilingual Whisper tiny model~\cite{whisper}.
Spatial fidelity is measured using angular and diffuseness errors computed between the reference and reconstructed signals.
The angular error compares the directions of the active-intensity vectors using a metric adapted from~\cite{foatokenizer}, while the diffuseness error is defined as the mean absolute error between the corresponding $D_{t,b}$ values.
Audio samples are available online\footnote{\url{https://weiting-lai.github.io/MetaFOA}}.

To evaluate whether the compressed representations retain task-relevant spatial information beyond reconstruction fidelity, we further assess the proposed method on downstream SELD, following the DCASE 2023 Challenge protocol on STARSS23~\cite{starss23}. 
We freeze both the RVQ codebooks of the proposed spatial metadata codec and the pretrained monaural signal codec, concatenate their latent representations, and train a lightweight probe, comprising three temporal convolutional layers and two fully connected layers, to predict Multi-ACCDOA~\cite{multi-accdoa} at a 100-ms label resolution.
We report the F-score and localization error (LE) with a $20^\circ$ spatial threshold.

\vspace{-0.3cm}
\subsection{FOA Reconstruction Results}
Table~\ref{tab:main_results} shows the FOA reconstruction performance on the three evaluation datasets.
With only 750~bps of spatial metadata, the proposed method remains competitive with the 48~kbps high-bitrate MPEG-H reference.
In contrast, while conventional DirAC performs well at 1.5~kbps, its performance degrades substantially at 750~bps, highlighting the difficulty of designing effective bit allocation under low bitrate constraints.
By separately coding the audio signal and spatial metadata, both the proposed method and DirAC remain robust when paired with different monaural signal codecs.
Moreover, because the $W$ channel is encoded independently by the selected signal codec, its content preservation, as reflected by WER, is determined by the signal codec only.
This allows these two methods to directly benefit from pretrained monaural codecs, resulting in more consistent WER performance than the end-to-end neural FOA codec.
Spatial errors are generally higher on STARSS23 and MEIR, where multi-source and reverberant conditions make spatial parameter estimation more challenging. This highlights the need for more reliable spatial metrics for evaluating reconstructed FOA signals in complex acoustic scenes.
Although we align the evaluation protocol and metrics as closely as possible, differences in experimental settings may remain, and the comparison should therefore not be interpreted as an absolute ranking between FOA-VQGAN and the other methods.


\vspace{-0.3cm}
\subsection{Effect of Parameter-Band Resolution}
We select a clip from STARSS23 to illustrate the reconstruction of the 3-D directivity vector at different parameter-band resolutions $B$, as shown in Fig.~\ref{fig:varyB}.
Under the same metadata bitrate of $0.75$~kbps, the two methods exhibit contrasting trends as $B$ increases.
For the proposed method, directivity distortion progressively decreases from $B=3$ to $36$, suggesting that finer parameter-band resolution better preserves frequency-dependent spatial variations.
This benefit comes without a bitrate penalty because the proposed representation decouples the spatial metadata rate from $B$ and jointly quantizes the cross-band 3-D directivity vectors. In contrast, increasing $B$ from $3$ to $5$ in conventional DirAC increases distortion, since the fixed bitrate must be distributed across more bands, forcing a coarser DOA grid $\mathcal{G}_{\Omega}$.
We did not further implement $B>5$ at $0.75$~kbps, as at $B=5$, the average DOA allocation is already reduced to 3 bits, corresponding to only eight directions. 
At the matched $B=3$ and $5$, the proposed method still yields lower distortion than DirAC, suggesting more efficient use of the limited codewords through joint cross-band quantization.

\begin{table}[t]
\centering
\caption{STARSS23 SELD results. Results from the uncompressed reference are underlined. Best results among codecs are in bold.}
\vspace{0.1cm}
\label{tab:seld}
\setlength{\tabcolsep}{5pt}
\renewcommand{\arraystretch}{1.0}
\begin{tabular}{llcc}
\toprule
\textbf{Method}
& \textbf{W-ch}
& \textbf{F-score $\uparrow$}
& \textbf{LE $\downarrow$} \\
\midrule
DCASE baseline~\cite{dcase2023baseline}
& -- & \underline{29.9} & \underline{$22^\circ$} \\

FOA-VQGAN~\cite{foatokenizer}
& -- & 11.1 & $37^\circ$ \\
\midrule

\multirow{2}{*}{Proposed}
& EnCodec & \textbf{14.2} & $\mathbf{30^\circ}$\\
& DAC & 13.8& $34^\circ$\\
\bottomrule
\end{tabular}
\vspace{-0.2cm}
\end{table}

\vspace{-0.2cm}
\subsection{Downstream SELD}
Table~\ref{tab:seld} shows the downstream SELD performance.
When paired with either EnCodec or DAC, the $k$-means-quantized spatial metadata achieves performance comparable to the end-to-end neural FOA codec, with EnCodec giving slightly better results.
While a gap remains when compared to the DCASE baseline trained on uncompressed continuous features, these results provide initial evidence that the proposed FOA codec can be flexibly paired with different monaural signal codecs while preserving key spatial cues for downstream tasks.

\vspace{-0.3cm}
\section{Conclusion}
\label{sec:conclusion}
In this paper, we proposed a constant-bitrate spatial metadata quantizer for a lightweight parametric FOA codec. DOA and diffuseness are combined into 3-D directivity vectors and jointly quantized across frequency bands using RVQ with codebooks fitted by stage-wise k-means. With only 750 bps of spatial metadata, the proposed method achieves competitive FOA reconstruction performance compared with conventional methods operating at higher bitrates, alongside downstream SELD performance comparable to FOA-VQGAN. Performance remains robust with either EnCodec or DAC as the monaural codec. Future work includes exploring spatial audio generation as a downstream task and extending the framework to higher-order Ambisonics.


\balance
\bibliographystyle{IEEEbib}
\bibliography{refs}

\end{document}